\documentclass[10pt]{extarticle}

\usepackage[utf8]{inputenc}
\usepackage[T1]{fontenc}
\usepackage{lmodern}
\usepackage[a4paper,margin=0.82in]{geometry}
\usepackage{setspace}
\usepackage{microtype}
\usepackage{hyperref}
\usepackage{graphicx}
\usepackage{array}
\usepackage{enumitem}
\usepackage{amsmath}
\usepackage{url}
\usepackage{float}
\usepackage{xcolor}
\usepackage{caption}
\usepackage{tabularx}
\usepackage{booktabs}
\usepackage{abstract}

\newcolumntype{Y}{>{\raggedright\arraybackslash}X}

\setlist[itemize]{leftmargin=1.4em, topsep=2pt, itemsep=2pt, parsep=0pt, partopsep=0pt}
\setlist[enumerate]{leftmargin=1.5em, topsep=2pt, itemsep=2pt, parsep=0pt, partopsep=0pt}

\hypersetup{
colorlinks=true,
linkcolor=blue,
citecolor=blue,
urlcolor=blue,
pdftitle={A Fuzzy Logic Framework for Community-Aware Crime Hotspot Detection},
pdfauthor={Ariton Verush and Vaibhav Motwani},
pdfkeywords={urban computing, fuzzy logic, crime hotspot detection, community reporting, predictive policing, data privacy, real-time notifications, human-centered systems}
}

\newcolumntype{L}[1]{>{\raggedright\arraybackslash}p{#1}}

\renewcommand{\arraystretch}{1.18}
\begin{document}

\begin{center}
\vspace{1.35cm}

{\fontsize{20}{24}\selectfont\bfseries
A Fuzzy Logic Framework for Community-Aware \\
Crime Hotspot Detection
\par}

\vspace{1.05cm}

{\large\itshape
Prototype Application Architecture and Exploratory Validation\\
for an Urban Computing Platform
\par}

\vspace{1.05cm}

{\normalsize
\textbf{Ariton Verush}\\
MSc in Computer Science\\
University of Bern\\
Bern, Switzerland\\
\texttt{ariton.verush@students.unibe.ch}
\par}

\vspace{0.55cm}

{\normalsize
\textbf{Vaibhav Motwani}\\
MSc in Computer Science\\
University of Bern\\
Bern, Switzerland\\
\texttt{vaibhav.motwani@students.unibe.ch}
\par}

\vspace{1.05cm}

{\normalsize
Urban Computing seminar report: December 2024\\
Revised and expanded as a framework paper: June 2026
\par}

\end{center}

\vspace{0.95cm}

\begin{abstract}
Urban crime prevention is a persistent socio-technical challenge for municipalities, law enforcement agencies, and citizens. Traditional reporting and response processes often rely on delayed incident reports and reactive resource allocation, while community-level signals and ambiguous early-warning indicators may remain underused. This paper reframes an Urban Computing seminar project into a fuzzy logic-based framework for community-aware urban crime hotspot detection and real-time notification. The proposed platform combines citizen reports, historical crime data, contextual urban indicators, and configurable fuzzy rules to estimate localized risk levels and support targeted awareness notifications. Unlike binary classification approaches, fuzzy logic can represent partial risk, uncertainty, and incomplete information, making it suitable for urban environments where risk is gradual and context-dependent. The paper presents the system architecture, fuzzy risk model, notification workflow, privacy safeguards, and exploratory validation with 25 participants. The evaluation focuses on perceived usefulness, notification relevance, usability, trust, privacy concerns, multilingual accessibility, and acceptance of community reporting. The findings suggest that such a platform may improve situational awareness, support reporting and planning discussions, and help address weak points in existing urban safety workflows, while not yet constituting evidence of deployed crime reduction or real-world predictive accuracy. The contribution is a responsible, human-centered framework for future research on fuzzy logic, urban computing, and community-aware crime prevention systems.
\end{abstract}


\newpage
\section{Introduction}

Urban crime prevention is not only a law enforcement problem; it is a socio-technical challenge involving citizens, institutions, urban planners, data infrastructures, and public trust. Cities contain dense and dynamic environments where theft, burglary, vandalism, parcel theft, and other property-related offenses may cluster around transport hubs, commercial areas, events, or neighborhoods with specific socio-spatial patterns \cite{Perry2013PredictivePolicing,Meijer2019PredictivePolicing,BFS2024Crime}. Traditional responses often depend on post-incident reporting, which can delay intervention and leave communities with limited situational awareness.

Data-driven and predictive policing systems have been proposed to improve crime forecasting, patrol planning, and resource allocation. However, these systems remain ethically sensitive. They may reproduce historical reporting biases, create feedback loops, or encourage over-policing if their outputs are treated as objective truth rather than decision support \cite{BennettMoses2018AlgorithmicPrediction,Goodson2021HotspotModels}. Therefore, any urban safety platform that uses prediction or risk classification should be designed with transparency, accountability, privacy protection, and human oversight.

This paper presents a fuzzy logic-based framework for community-aware urban crime hotspot detection and real-time notification. The work builds on an Urban Computing seminar project that proposed a two-part platform: a citizen notification system and a crime hotspot detection system. The revised paper reframes that project as a responsible framework and exploratory validation study rather than as a fully deployed predictive policing system.

Fuzzy logic is appropriate for this context because urban crime risk is rarely binary. A location is not simply safe or dangerous. Risk may be minimal, low, relatively low, moderate, high, or very high depending on historical incidents, crowd density, time of day, public events, recent reports, and contextual conditions. Fuzzy logic allows these conditions to be represented as degrees of membership rather than hard categories \cite{Zadeh1965FuzzySets,Mamdani1975FuzzyControl}.

The framework has two core components. First, a fuzzy hotspot detection component estimates localized risk levels from historical, contextual, and community-generated signals. Second, a real-time notification component communicates relevant alerts while avoiding unnecessary panic, over-notification, and privacy leakage. The goal is not to automate policing decisions but to support earlier awareness, better reporting channels, and more informed resource-planning discussions.

\subsection{Research Questions}

The paper is guided by four research questions:

\begin{itemize}
\item \textbf{RQ1:} How can fuzzy logic model ambiguous and incomplete urban crime-risk indicators in a community-aware crime prevention platform?
\item \textbf{RQ2:} How can real-time notifications and community reporting be integrated without producing unnecessary panic, privacy risks, or over-surveillance?
\item \textbf{RQ3:} How do participants perceive the usefulness, usability, trustworthiness, privacy safeguards, and accessibility of a crime hotspot and notification prototype?
\item \textbf{RQ4:} What can an exploratory validation study support, and what claims must remain outside the scope of preliminary evaluation?
\end{itemize}

\subsection{Contributions of the Paper}

This paper makes five main contributions:

\begin{itemize}
\item it reframes a seminar-based urban safety prototype into a structured framework paper for community-aware crime hotspot detection;
\item it defines a fuzzy risk model that represents urban safety as gradual and context-dependent rather than binary;
\item it presents a modular architecture connecting report collection, preprocessing, fuzzy inference, geospatial analysis, notifications, and governance safeguards;
\item it reports an exploratory validation with 25 participants, focusing on perceived usefulness, notification relevance, usability, privacy, trust, multilingual accessibility, and community-reporting acceptance;
\item it documents a Python-based prototype implementation and map-based interface used to instantiate the proposed framework.
\end{itemize}

\section{Background and Related Work}

\subsection{Predictive Policing and Crime Hotspot Detection}

Predictive policing refers to the use of statistical, algorithmic, or computational methods to support decisions about where crime may occur, who may be at risk, or how law enforcement resources may be allocated \cite{Perry2013PredictivePolicing}. Crime hotspot detection is one of the most common applications, focusing on areas with recurrent or elevated incident patterns \cite{Braga2019HotSpots,Hu2018STKDE,Goodson2021HotspotModels}.

Recent reviews of AI-based crime prediction also show that crime prediction research spans statistical models, machine learning, and hybrid decision-support systems, but that evaluation quality, transparency, and deployment context remain central concerns \cite{Dakalbab2022AIcrime}.

Prior work has shown that predictive policing can support planning, but it must be interpreted carefully. Meijer and Wessels review benefits and drawbacks, emphasizing that predictive policing may support efficiency but also raises questions of accountability, legitimacy, bias, and privacy \cite{Meijer2019PredictivePolicing}. Bennett Moses and Chan similarly argue that algorithmic policing requires careful attention to assumptions, evaluation, and accountability \cite{BennettMoses2018AlgorithmicPrediction}.

Hotspot policing and predictive enforcement are not equivalent to crime prevention by themselves. Even when a model identifies a high-risk location, crime reduction depends on how institutions respond, how communities perceive the intervention, and whether the system avoids reinforcing historical inequalities \cite{Braga2019HotSpots,Goodson2021HotspotModels,Adensamer2019PredictivePolicing,Joh2017FedUp}. This paper therefore treats prediction as decision support, not as automated enforcement.

\subsection{Fuzzy Logic and Interpretable Risk Modeling}

Fuzzy logic was introduced by Zadeh as a way to reason with gradual membership rather than crisp true/false categories \cite{Zadeh1965FuzzySets}. Mamdani and Assilian later demonstrated fuzzy control as an interpretable rule-based approach for systems where linguistic reasoning and uncertainty are important \cite{Mamdani1975FuzzyControl}.

Urban safety is a natural candidate for fuzzy reasoning because many relevant variables are imprecise. Crowd density may be low, moderate, or high. Crime history may be limited, typical, or elevated. Public-event traffic may be normal or exceptional. A fuzzy inference system can combine these variables into a risk score while preserving interpretable rules.

Compared with black-box predictive systems, fuzzy rules are easier to inspect and revise. For example, a rule such as ``if crowd density is high and recent theft reports are high, then risk is high'' can be discussed with domain experts and community stakeholders. This interpretability is important in public-sector and policing-related contexts, where explainability and accountability matter.

\subsection{Community Reporting and Human-Centered Urban Computing}

Community reporting systems allow citizens to submit observations, incident reports, or safety concerns. These reports can supplement official data, especially when incidents are underreported or when residents notice recurring local patterns. However, community reporting introduces challenges such as false reports, privacy concerns, unequal participation, and potential stigmatization of neighborhoods.

Human-centered urban computing should therefore treat citizens as stakeholders rather than passive data sources. A community-aware platform must provide understandable notifications, consent mechanisms, privacy controls, multilingual access, feedback channels, and transparent limitations. In this paper, community reporting is integrated as one signal among several, not as an unchecked source of enforcement action.

\subsection{Privacy, Law, and Responsible AI}

Crime-related data can be sensitive even when it does not contain direct identifiers, especially when location traces, alerts, or predictive-risk labels create surveillance-like effects \cite{vanBrakel2016Surveillance}. Location data, incident narratives, and community reports may still reveal personal patterns or indirectly identify individuals. In Switzerland, the Federal Act on Data Protection provides a legal framework for the processing of personal data, including principles such as proportionality, purpose limitation, and data minimization \cite{FADP2023}. Broader European discussions around AI regulation also emphasize risk management, transparency, and safeguards for high-risk systems \cite{EUAIAct2024}.

For this reason, the proposed framework includes privacy-by-design and governance requirements. These include anonymization, role-based access control, audit logging, retention limits, transparency reports, and human review. The system is not intended to replace law enforcement judgment or automate coercive decisions.

\section{Methodology}

This paper is based on a seminar project in Urban Computing and a later revision into a framework-style preprint. The original project proposed a fuzzy logic-based platform for urban crime prevention, combining real-time notifications, hotspot detection, public feedback, and police resource-planning support. The present version restructures the project into a more academically cautious paper with citations, clearer scope, and explicit limitations.

The methodology has four parts:

\begin{enumerate}
\item \textbf{Problem analysis:} identifying pain points in urban crime prevention, including delayed reporting, resource allocation challenges, public awareness gaps, privacy risks, and community trust.
\item \textbf{Framework design:} defining a modular architecture for data collection, preprocessing, fuzzy risk inference, geospatial analysis, notifications, storage, and user interfaces.
\item \textbf{Fuzzy model construction:} formalizing input variables, linguistic categories, rules, risk outputs, and notification levels.
\item \textbf{Exploratory validation:} evaluating the prototype concept with 25 participants, focusing on perceived usefulness, usability, trust, privacy, notification relevance, and community-reporting acceptance.
\end{enumerate}

The validation is exploratory rather than conclusive. It can support claims about user perception, design feasibility, and acceptance of the concept. It cannot fully validate actual crime reduction, deployed predictive accuracy, fairness across demographic groups, long-term public trust, or operational effects on police workload. Those outcomes require larger datasets, field deployment, institutional collaboration, and longitudinal evaluation.

\section{System Framework}

\subsection{Design Goals}

The proposed platform is designed around five goals:

\begin{itemize}
\item \textbf{Earlier awareness:} notify citizens about relevant local safety risks before they become personally affected.
\item \textbf{Community participation:} allow citizens to contribute observations and reports in a controlled and privacy-aware way.
\item \textbf{Interpretable risk modeling:} use fuzzy rules that can be inspected and revised by stakeholders.
\item \textbf{Responsible notification:} prevent panic, over-notification, and stigmatization by filtering alerts according to risk level and relevance.
\item \textbf{Decision support:} support planning discussions without automating enforcement or replacing human judgment.
\end{itemize}

\subsection{Platform Architecture}

The platform is organized as a pipeline connecting data sources, risk modeling, storage, and user-facing interfaces. Table~\ref{tab:architecture} summarizes the architecture in a compact format.

\begin{table}[H]
\centering
\caption{Core architecture of the proposed crime hotspot and notification framework}
\label{tab:architecture}
\small
\renewcommand{\arraystretch}{1.2}
\begin{tabularx}{\textwidth}{L{3.2cm}Y}
\toprule
\textbf{Layer} & \textbf{Role in the system} \\
\midrule
\textbf{Data collection} &
Collects historical crime data, victim reports, optional citizen feedback, event information, and contextual urban indicators. \\
\addlinespace
\textbf{Preprocessing} &
Cleans reports, normalizes variables, geocodes locations, extracts temporal features, and prepares data for fuzzy inference. \\
\addlinespace
\textbf{Fuzzy risk inference} &
Combines linguistic variables and fuzzy rules to estimate localized risk levels for hotspot detection and notification decisions. \\
\addlinespace
\textbf{Geospatial analysis} &
Maps reports and risk scores to urban zones, supports geo-fencing, and enables hotspot visualization for administrators. \\
\addlinespace
\textbf{Notification system} &
Filters alerts by risk level, location, user preferences, and relevance in order to reduce panic and notification fatigue. \\
\addlinespace
\textbf{Governance and storage} &
Stores reports, notification logs, audit records, and model outputs under retention limits and role-based access control. \\
\bottomrule
\end{tabularx}
\end{table}

\subsection{Prototype Implementation and Map Interface}

In addition to the conceptual framework, the seminar project included a Python-based prototype implementation for crime hotspot visualization and risk communication. The prototype demonstrates how localized crime-risk categories can be displayed on an interactive map and how fuzzy risk levels may be translated into user-facing hotspot indicators. The implementation was developed as a proof of concept rather than as a deployed municipal system. Therefore, the prototype is used in this paper to illustrate feasibility, interface direction, and framework instantiation, not to claim operational predictive accuracy.

\begin{figure}[H]
\centering
\includegraphics[width=0.94\textwidth]{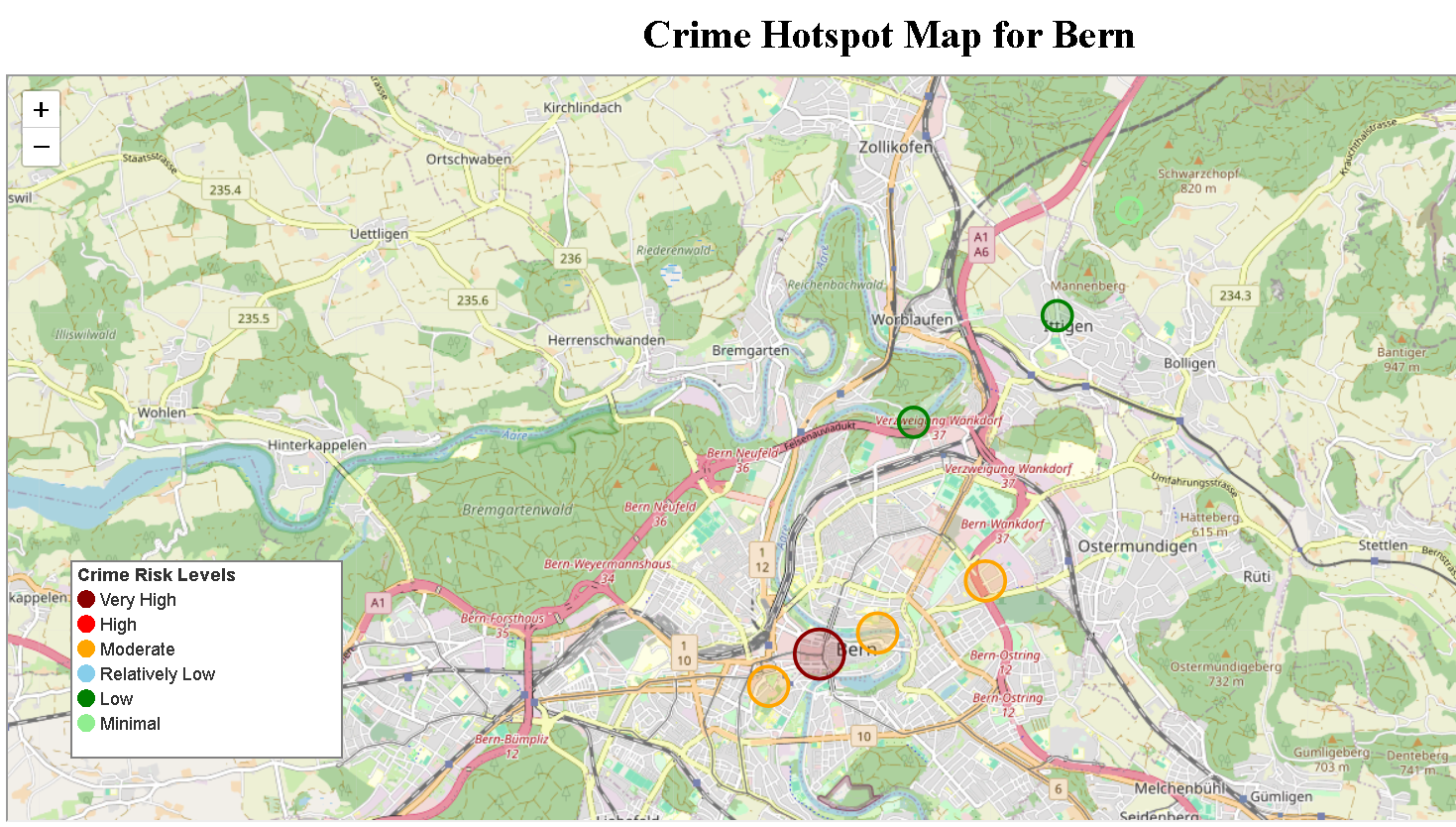}
\caption{Prototype crime hotspot map for Bern, illustrating color-coded localized risk levels used in the proposed framework. The figure is an illustrative prototype interface and not evidence of deployed predictive accuracy. Map data: \copyright{} OpenStreetMap contributors.}
\label{fig:bern-hotspot-map}
\end{figure}

Figure~\ref{fig:bern-hotspot-map} illustrates the prototype direction of the crime hotspot map interface. The visualization shows how localized risk levels can be communicated spatially through color-coded urban zones.

\subsection{User Roles}

The framework distinguishes between citizens, administrators, and institutional stakeholders. Citizens receive localized safety notifications, submit optional reports, and configure notification preferences. Administrators monitor hotspots, review reports, and manage alerts. Institutional stakeholders such as municipal authorities or police departments may use aggregate risk trends for planning, but should not treat the system as an automatic enforcement tool.

\subsection{Notification Workflow}

Notifications are generated only after risk levels are interpreted through contextual and governance filters. For example, a moderate risk score may trigger a general awareness message, while a very high score may require administrative review before a public alert. This prevents risk scores from being translated directly into public fear or coercive action.

The notification workflow follows four steps:

\begin{enumerate}
\item collect and preprocess new reports or contextual signals;
\item update the fuzzy risk level for the relevant zone;
\item apply notification rules based on severity, location, frequency, and user preferences;
\item send an appropriate alert or route the case for administrative review.
\end{enumerate}

\section{Fuzzy Risk Model}

\subsection{Input Variables}

The fuzzy model estimates localized crime risk from multiple uncertain variables. Table~\ref{tab:fuzzy-inputs} summarizes the initial input variables used to instantiate the rule base. These variables are not intended to be exhaustive; they define the first version of a transparent rule base that can be extended with domain expertise.

\begin{table}[H]
\centering
\caption{Input variables for the fuzzy risk model}
\label{tab:fuzzy-inputs}
\small
\renewcommand{\arraystretch}{1.2}
\begin{tabularx}{\textwidth}{L{3.2cm}L{3.1cm}Y}
\toprule
\textbf{Input variable} & \textbf{Example fuzzy labels} & \textbf{Interpretation} \\
\midrule
Historical crime rate & Low, moderate, high & Frequency of relevant past incidents in an urban zone. \\
Crowd density & Sparse, normal, crowded & Approximate public presence based on time, events, transport flow, or mobility patterns. \\
Recent reports & Few, increasing, many & Short-term increase in official reports or community-submitted observations. \\
Event traffic & Normal, elevated, exceptional & Temporary changes caused by festivals, protests, sports events, nightlife, or holidays. \\
Time context & Daytime, evening, night & Temporal conditions that may influence risk patterns and notification relevance. \\
Location sensitivity & Ordinary, sensitive, critical & Places such as schools, hospitals, transport hubs, public institutions, or commercial zones. \\
\bottomrule
\end{tabularx}
\end{table}

\subsection{Membership and Risk Output}

Each input is represented through fuzzy membership functions. For a given urban zone \(z\), the system evaluates variables such as historical crime rate \(C_z\), crowd density \(D_z\), recent reports \(R_z\), event traffic \(E_z\), and temporal context \(T_z\). The fuzzy inference process can be expressed abstractly as:

\[
Risk_z = F(C_z, D_z, R_z, E_z, T_z)
\]

where \(F\) denotes a rule-based fuzzy inference function. After inference and defuzzification, the system maps the resulting value to one of six risk levels, as summarized in Table~\ref{tab:risk-levels}. These levels are design categories rather than universal thresholds; in a real deployment, they would require local calibration, expert review, and fairness auditing \cite{Zadeh1965FuzzySets,Mamdani1975FuzzyControl,BennettMoses2018AlgorithmicPrediction}.

\begin{table}[H]
\centering
\caption{Risk levels and notification interpretation}
\label{tab:risk-levels}
\small
\renewcommand{\arraystretch}{1.2}
\begin{tabularx}{\textwidth}{L{2.5cm}L{2.0cm}Y}
\toprule
\textbf{Risk level} & \textbf{Score range} & \textbf{Notification interpretation} \\
\midrule
Minimal & 0--15 & No public alert; normal monitoring only. \\
Low & 16--30 & No urgent alert; optional safety reminder if relevant. \\
Relatively low & 31--45 & Soft contextual notification for users who opted into low-intensity alerts. \\
Moderate & 46--60 & Local awareness message; wording must avoid panic and indicate uncertainty. \\
High & 61--80 & Stronger alert or administrator review depending on report verification and context. \\
Very high & 81--100 & Administrative review before public notification; possible resource-planning escalation. \\
\bottomrule
\end{tabularx}
\end{table}

The score ranges are design parameters rather than universal thresholds. In a real deployment, they would need calibration with local data, expert feedback, and fairness evaluation.

\subsection{Rule Base}

The rule base makes risk reasoning inspectable. Table~\ref{tab:fuzzy-rules} summarizes example fuzzy rules used for hotspot detection and notification.

\begin{table}[H]
\centering
\caption{Example fuzzy rules for hotspot detection and notification}
\label{tab:fuzzy-rules}
\small
\renewcommand{\arraystretch}{1.22}
\begin{tabularx}{\textwidth}{p{1.1cm}X}
\toprule
\textbf{ID} & \textbf{Rule and intended response} \\
\midrule
\textbf{R1} &
If historical crime rate is high and crowd density is high, then risk is very high. This rule detects crowded high-risk zones where situational awareness may be needed. \\
\addlinespace
\textbf{R2} &
If historical crime rate is moderate and crowd density is sparse, then risk is low. This prevents unnecessary alerting in low-activity areas. \\
\addlinespace
\textbf{R3} &
If recent theft reports are increasing and the time context is evening, then risk is high. This captures short-term theft patterns that may require localized awareness. \\
\addlinespace
\textbf{R4} &
If event traffic is exceptional and historical crime rate is moderate, then risk becomes moderate or high. This handles temporary public events without permanently labeling an area as dangerous. \\
\addlinespace
\textbf{R5} &
If parcel theft reports increase during holidays, then nearby residents receive a soft prevention-oriented alert. This supports seasonal awareness without creating panic. \\
\addlinespace
\textbf{R6} &
If location sensitivity is critical and recent reports are many, then the case is routed to administrator review before public notification. This protects sensitive places and avoids automatic escalation. \\
\addlinespace
\textbf{R7} &
If risk is high but reports are unverified, then the system delays public alerting and requests verification. This reduces false alarms and protects public trust. \\
\bottomrule
\end{tabularx}
\end{table}

This rule base is intentionally transparent. It allows municipal stakeholders, law enforcement, and community representatives to inspect why a risk category was produced and how the system should respond.

\section{Exploratory Validation Study}

\subsection{Participants and Procedure}

The revised evaluation includes 25 participants. The purpose was to evaluate whether the concept is understandable, useful, trustworthy, and acceptable as an early-stage urban computing prototype. Participants were introduced to the platform concept, including hotspot visualization, fuzzy risk levels, real-time notifications, community reporting, and privacy safeguards.

The evaluation focused on perception and design feasibility. It did not measure deployed crime reduction or real-world predictive accuracy. Instead, it assessed whether participants believed the system could improve awareness, reporting, communication, and planning if implemented responsibly.

\subsection{Evaluation Dimensions}

The evaluation measured six dimensions:

\begin{itemize}
\item perceived usefulness for awareness and prevention support;
\item relevance and clarity of localized notifications;
\item usability and interface understandability;
\item trust and privacy acceptance under safeguards;
\item acceptance of community reporting;
\item need for multilingual and accessible design.
\end{itemize}

Table~\ref{tab:validation-results} summarizes the extended 25-participant evaluation. The values are reported as rounded aggregate results from the revised seminar evaluation.

\begin{table}[H]
\centering
\caption{Exploratory validation results with 25 participants}
\label{tab:validation-results}
\small
\renewcommand{\arraystretch}{1.22}
\begin{tabularx}{\textwidth}{p{3.3cm}p{3.2cm}Y}
\toprule
\textbf{Evaluation criterion} & \textbf{Result} & \textbf{Interpretation} \\
\midrule

\textbf{Perceived usefulness} &
22/25 positive (88\%); 2/25 neutral (8\%) &
Most participants believed the system could support awareness and prevention-oriented behavior. \\
\addlinespace

\textbf{Notification relevance} &
21/25 positive (84\%); 3/25 neutral (12\%) &
Localized alerts were seen as useful if they remain relevant, moderate, and non-sensational. \\
\addlinespace

\textbf{Usability and clarity} &
19/25 positive (76\%); 4/25 neutral (16\%) &
Participants generally understood the workflow, but requested clearer explanations of risk categories. \\
\addlinespace

\textbf{Trust and privacy safeguards} &
18/25 positive (72\%); 5/25 neutral (20\%) &
Participants accepted the concept when anonymization, encryption, and human oversight were emphasized. \\
\addlinespace

\textbf{Community reporting acceptance} &
23/25 positive (92\%); 1/25 neutral (4\%) &
Anonymous or protected reporting was viewed positively, especially for recurring local problems. \\
\addlinespace

\textbf{Multilingual and accessibility need} &
19/25 positive (76\%); 5/25 neutral (20\%) &
Participants supported multilingual and accessible design as important for broader adoption. \\

\bottomrule
\end{tabularx}
\end{table}

\subsection{Scenario-Level Findings}

Participants were also asked to reason about hypothetical hotspot and notification scenarios. Around 20 of 25 participants interpreted the fuzzy risk categories consistently with the expected scenario logic. This supports the clarity of the risk model at a conceptual level, but it should not be interpreted as predictive accuracy. The result shows that the categories were understandable to participants, not that the model predicts real crime events.

Participants especially valued three aspects: localized alerts, the possibility of anonymous community reports, and the use of risk categories rather than binary safe/dangerous labels. At the same time, they raised concerns about excessive alerts, privacy, false reports, and the risk of stigmatizing neighborhoods.

\subsection{Qualitative Feedback}

The qualitative feedback suggests that the system should be designed as an awareness and reporting tool rather than as a fear-generating crime map. Participants preferred alerts that explain uncertainty and provide practical guidance. Several participants also emphasized that reports should be verified or moderated before being used in public alerts.

\newpage
The most common improvement requests were:

\begin{itemize}
\item clearer explanations of risk levels;
\item multilingual interface support;
\item stronger data protection and two-factor authentication;
\item anonymous community reporting with moderation;
\item avoidance of neighborhood stigmatization;
\item clear separation between confirmed incidents and predicted risk.
\end{itemize}

\section{Privacy, Ethics, and Governance}

\subsection{Privacy-by-Design}

A crime reporting and hotspot detection platform must be privacy-preserving from the beginning. The framework therefore follows six privacy-by-design principles:

\begin{itemize}
\item \textbf{Purpose limitation:} data is collected only for awareness, reporting, analysis, and planning support.
\item \textbf{Data minimization:} the system collects only the fields necessary for risk estimation and notification.
\item \textbf{Anonymization and pseudonymization:} personal identifiers are removed or separated from analytical records wherever possible.
\item \textbf{Access control:} sensitive reports are available only to authorized roles.
\item \textbf{Retention limits:} outdated reports and logs are deleted according to defined policies.
\item \textbf{Transparency:} users can see what types of data are collected and how they influence the system.
\end{itemize}

\subsection{Bias and Fairness Risks}

Predictive policing systems can reproduce bias if historical data reflects unequal reporting, unequal patrol patterns, or socio-economic inequalities \cite{BennettMoses2018AlgorithmicPrediction,Goodson2021HotspotModels}. Therefore, the framework avoids treating risk scores as automatic enforcement commands. Human review, community oversight, and transparent reporting are essential.

The fuzzy model also needs periodic auditing. If one neighborhood is repeatedly marked as high risk, the system should examine whether the pattern reflects actual incidents, reporting bias, demographic bias, or system configuration. Without such safeguards, even interpretable systems can contribute to unfair outcomes.

\subsection{Responsible Notification}

Notifications should inform rather than alarm. A responsible alert should be localized, relevant, and actionable. It should avoid sensational language and clearly distinguish between confirmed incidents and risk estimates. Users should also be able to control notification intensity.

Responsible notification design includes:

\begin{itemize}
\item configurable alert frequency;
\item category-based notification preferences;
\item soft wording for moderate or uncertain risks;
\item administrative review for high-risk public alerts;
\item visible explanation of why an alert was sent.
\end{itemize}

\section{Comparison with Existing Approaches}

The proposed framework combines ideas from community-reporting apps, predictive hotspot systems, and privacy-aware decision support. Existing approaches often emphasize one dimension more strongly than the others. Citizen-style applications emphasize public alerts and reporting, as reflected by commercial community-alert systems such as Citizen \cite{CitizenOfficial}. Predictive hotspot systems emphasize data-driven resource allocation. Police intelligence systems may focus on organized crime and internal analysis. The proposed framework instead tries to combine fuzzy risk modeling, community participation, privacy safeguards, and responsible notification. Table~\ref{tab:comparison} summarizes this conceptual positioning.

\begin{table}[H]
\centering
\caption{Conceptual comparison of urban crime prevention approaches}
\label{tab:comparison}
\small
\renewcommand{\arraystretch}{1.18}
\begin{tabularx}{\textwidth}{p{3.3cm}p{2.3cm}p{2.5cm}Y}
\toprule
\textbf{Feature} & \textbf{Community alert apps} & \textbf{Predictive hotspot systems} & \textbf{Proposed fuzzy framework} \\
\midrule
Public notifications & Strong & Usually limited & Strong, but filtered by relevance and risk level. \\
Predictive modeling & Limited & Strong & Moderate, interpretable, and fuzzy-rule based. \\
Community reporting & Strong & Usually limited & Strong, with moderation and privacy safeguards. \\
Interpretability & Variable & Variable & High, because rules can be inspected. \\
Privacy governance & Variable & Often internal & Explicit design requirement. \\
Human-centered design & Strong for citizens & Often institution-centered & Balances citizens, administrators, and institutions. \\
\bottomrule
\end{tabularx}
\end{table}

This comparison is conceptual, not a performance ranking. Its purpose is to clarify the design position of the proposed framework. The framework should be judged by whether it supports awareness, responsible reporting, interpretable risk modeling, and accountable governance.

\section{Discussion}

\subsection{What the Validation Supports}

The 25-participant study supports the platform as an exploratory urban computing concept. Participants found the idea useful for awareness, reporting, and localized notification. They also identified concrete design requirements, including privacy safeguards, multilingual access, moderation, and clear risk explanations.

This is meaningful because public trust and usability are central to any safety-related technology. A technically strong system that citizens do not trust or understand would be difficult to deploy responsibly. The evaluation therefore validates early-stage acceptance and design feasibility.

\subsection{What the Validation Does Not Prove}

The study does not prove actual crime reduction, deployed predictive accuracy, long-term trust, fairness across demographic groups, or operational impact on police workload. Those outcomes require real deployment, larger datasets, institutional cooperation, and longitudinal analysis.

However, this limitation does not eliminate the value of the study. Early validation can identify whether a concept is understandable and acceptable before it is deployed. It can also reveal ethical risks earlier, which is especially important for policing-related technologies.

\subsection{Preventive Potential}

The framework may still help reduce or prevent some problems indirectly. It may improve public awareness, strengthen reporting channels, identify recurring weak areas, and support better planning discussions. For example, localized alerts can remind citizens to protect belongings in theft-prone zones, while community reports can reveal patterns that may not appear immediately in official data.

The correct framing is therefore preventive support rather than automatic prevention. The platform may help improve weak areas in urban safety workflows, but it must remain transparent, proportionate, and accountable.

\newpage
\section{Limitations}

This paper has several limitations. First, the evaluation is exploratory and based on 25 participants. This is sufficient for preliminary validation of perceived usefulness and acceptability, but not for full empirical validation of crime-prevention outcomes, which would require field-style evaluation comparable to prior predictive-policing experiments \cite{Saunders2016Chicago,Hunt2014Shreveport}.

Second, the prototype has not been deployed with live municipal crime databases. The fuzzy risk model is therefore a framework-level contribution rather than a fully calibrated operational model.

Third, the evaluation relies on participant perceptions and hypothetical scenarios. Such feedback is valuable for design, but it cannot replace field testing.

Fourth, predictive policing and hotspot systems carry ethical risks. Even interpretable fuzzy models can produce harmful outcomes if deployed without auditing, community oversight, and legal safeguards.

Finally, the exact thresholds, rule weights, and notification policies require local adaptation. A model suitable for one city may not transfer directly to another.

\section{Future Work}

Future work should proceed in four directions. First, the fuzzy rule base should be refined with domain experts, urban planners, community representatives, and law enforcement stakeholders. Second, the prototype should be tested using anonymized historical crime data to evaluate whether fuzzy risk categories align with known hotspot patterns. Third, the notification system should be evaluated for alert fatigue, comprehension, and trust over longer periods.

Fourth, future work should include fairness and bias analysis. The system should be audited to ensure that risk classifications do not disproportionately stigmatize specific neighborhoods or groups. Transparency reports, privacy impact assessments, and public consultation should be part of any deployment plan.

A future larger study could compare three versions of the platform: one with no prediction, one with traditional hotspot scoring, and one with fuzzy risk explanation. This would allow researchers to test whether fuzzy explanations improve trust, understanding, and responsible use.

\section{Conclusion}

This paper presented a fuzzy logic-based framework for community-aware urban crime hotspot detection and real-time notification. The framework combines citizen reports, contextual indicators, fuzzy risk modeling, geospatial analysis, privacy safeguards, and notification design.

The exploratory validation with 25 participants suggests that the platform concept is perceived as useful and understandable, especially for localized awareness, community reporting, and prevention-oriented communication. Participants also emphasized privacy, trust, multilingual access, and the need to avoid fear-inducing notifications.

The paper does not claim full validation of crime reduction or deployed predictive accuracy. Instead, it contributes a responsible framework for future research on human-centered urban crime prevention. The central argument is that urban safety technologies should not be designed only as predictive engines. They should be transparent, community-aware, privacy-preserving, and accountable decision-support systems.

\section*{Code Availability}

A Python-based prototype implementation was developed as part of the Urban Computing seminar project. The prototype code for the crime hotspot identifier is available at:

\begin{center}
\url{https://github.com/Ariton123/engineering-lab/tree/main/coursework/master/seminar-urban-computing/crime-hotspot-identifier}
\end{center}

The repository is provided for transparency and illustration of the proposed framework. It should be interpreted as a proof-of-concept implementation, not as a validated operational crime prediction or law enforcement deployment system.

\end{document}